\documentclass[a4paper,fleqn]{cas-sc}

\usepackage[numbers]{natbib}

\def\tsc#1{\csdef{#1}{\textsc{\lowercase{#1}}\xspace}}
\tsc{WGM}
\tsc{QE}

\newcommand{\D}{{\rm d}}
\newcommand{\bd}{{\rm D}}
\renewcommand\({\left(} 
\renewcommand\){\right)}
\renewcommand\[{\left[} 
\renewcommand\]{\right]}

\begin{document}
\let\WriteBookmarks\relax
\def\floatpagepagefraction{1}
\def\textpagefraction{.001}

\shorttitle{}    

\shortauthors{}  

\title [mode = title]{Relativity from the Perspectives of Observers}  



%

\author[1]{Tao Wang}

\fnmark[1]

\ead{wangtao97@ustc.edu.cn}



\affiliation[1]{organization={Department of History of Science and Scientific Archaeology, University of Science and Technology of China},
            city={Hefei},
            postcode={230026}, 
            country={China}}

\author[2,3]{Yu Shi}

\cormark[1]

\fnmark[2]

\ead{yu_shi@ustc.edu.cn}



\affiliation[2]{organization={Wilczek Quantum Center, Shanghai Institute for Advanced Studies USTC},
            city={Shanghai},
            postcode={201315}, 
            country={China}}
               
\affiliation[3]{organization={University of Science and Technology of China},
            city={Hefei},
            postcode={230026}, 
            country={China}}

\cortext[1]{Corresponding author}



\begin{abstract}
This paper reviews the role of observers in the development of relativity theory, from special relativity to general relativity, emphasizing that observer‑dependent descriptions are as fundamental as the covariance of physical laws. After the introduction of a geometric framework for observers  using timelike worldlines, Frenet‑Serret formulas, projection operators, and the Frobenius condition for hypersurface‑orthogonal families, the paper revisits key problems in early relativistic mechanics, such as  the transformation of velocity and acceleration, the variational principle for particle motion, and the Ehrenfest paradox concerning rigid rotation. It shows that while early physicists often conflated coordinate systems with reference frames, their results remain valid because the underlying geometric objects are observer‑independent. The historical analysis, from Einstein’s 1905 work to the development of general relativity and later advances such as Hawking radiation, demonstrates that clarifying the concept of observers not only resolved paradoxes but also paved the way toward a field‑theoretic formulation of gravity. The paper concludes that observer dependence, far from being a nuisance, is an essential ingredient for understanding spacetime physics.
\end{abstract}


\begin{highlights}
\item Clarifies why early coordinate-frame conflations yielded valid results.
\item Uses variational methods to derive observer-dependent equations of motion
\item Reveals how Einstein's shift to action principles led to General Relativity
\end{highlights}


\maketitle

\section{Introduction}
\label{intro}

Special relativity (SR) emerged from the investigation of the property of transformation  of electrodynamics between different reference frames. Since then, the principle that the form of a physical law should remain covariant under coordinate transformation has become deeply ingrained in the minds of physicists. While the covariance of physical laws under coordinate transformation is of great importance, another crucial aspect of relativity, {\it i.e.}, the observer-dependence of physical phenomena, should not be overlooked. Experimental tests of relativity are all based on the observation data, thus a systematic approach to understanding and handling  them  is essential. 

Observation issues have consistently played a significant role in the development of relativity, repeatedly influencing its progression. Initially, Einstein examined how length, time interval, velocity, and acceleration transform between different reference frames\cite{einstein1905elektrodynamik}. Later, Minkowski introduced his `{\it world postulate}'\cite{minkowski1909}, to emphasize the geometric nature of the spacetime. He pointed out that different experimental data obtained in various reference frames are merely projections of the same geometric objects onto those frames. Such a visionary idea, in fact, has permeated the entire subsequent development of relativistic mechanics, even in the context of general relativity (GR).

Although SR  centers on reference frames and observers, the advent of GR turned the focus toward the covariance of physical laws. But in fact, the reference frame and observers also have their corresponding description in GR. This paper uses the description of observers in GR to look back, discussing  several issues in early stage of SR, at a vantage   point of view to see why some conclusions in SR were  correct even when this theory had some ambiguity. It is necessary to review the original ideas of establishing a theory at the early stage from a more complete view of point. In this way, we can clarify the relation between GR and SR and understand the status of observers in relativity.

This paper is organized as follows. In Section~\ref{sec2}, we review the framework for describing observers, establishing consistent notations for later discussions. Section~\ref{sec3} revisits several interesting problems in early relativistic mechanics using the geometric description of observers. Section~\ref{sec4} reviews the historical context of the theory's formative period and reassesses the conceptual standing of these entities. Finally, Section~\ref{sec5} provides an overall summary. Throughout the  paper, we use  metric signature $(-,+,+,+)$  and natural units with $c = \hbar = G = k_{B} = 1$.

\section{Description of observers}
\label{sec2}

This section aims to elucidate the concept of observers and introduce relevant notations. We   focus   on a broadly applicable definition of observers, rather than providing an exhaustive overview of existing work \cite{ellis1971relativistic,misner1973gravitation,sachs1977general,
thorne1982electrodynamics,ellis1999cosmological,
de2010classical,gourgoulhon20123}. This specific clarification provides the foundational clarity necessary for a more rigorous comprehension of historical discussions.

\subsection{A single observer and a family of observers}

It is well known that one of the most significant distinctions of relativity from the Newtonian framework is that `simultaneity' is no longer an absolute concept, which makes `synchronization' fundamentally important. From a geometric perspective, this issue is essentially a discussion about the reasonable foliations permitted by spacetime. To delve into the essence, we begin with the description of a single observer.

The evolution of an observer can be represented in terms of a timelike worldline, therefore a precise characterization of a curve in spacetime is necessary. The Frenet-Serret formulas, originally developed for curves in Euclidean space,  provide such a description and can be extended to curved spacetime\cite{synge1960gr}. These formulas describe the evolution of the unit tangent vector, normal, and binormal vectors of a given curve. In 3-dimensional Euclidean space, the tangent space at any point can be spanned by three linearly independent vectors, making the orthonormal basis consisting of tangent, normal, and binormal vectors a natural choice for a local frame. The choice of a local frame is not unique. A more general formalism to characterize the motion of local frame is the method of moving frame, introduced by E.~Cartan\cite{chen1999lectures}. However, as our focus is not on such generality, the Frenet-Serret formalism suffices for the present discussion.

If we want to generalize the Frenet-Serret formulas to GR, one of the changes arises from the fact that the spacetime signature differs from that of Euclidean space. Furthermore, because the analysis involves vectors that are transported along a curve in the curved spacetime under consideration, one must take into account the affine connection, which governs the comparison of vectors at different points. The natural generalization of Frenet-Serret formulas in spacetime would be
\begin{align}\label{FS-formula-GR}
	\dfrac{\bd}{\D \tau}\begin{pmatrix}
		U^{\mu}\\
		\hat{A}^{\mu}\\
		\hat{B}^{\mu}\\
		\hat{C}^{\mu}
	\end{pmatrix}=
	\begin{pmatrix}
		0&\alpha &0&0\\
		\alpha &0&\delta &0\\
		0&-\delta &0&\eta\\
		0&0&-\eta &0
	\end{pmatrix}
		\begin{pmatrix}
		U^{\mu}\\
		\hat{A}^{\mu}\\
		\hat{B}^{\mu}\\
		\hat{C}^{\mu}
	\end{pmatrix}
\end{align}
where $\bd/\D \tau$ represents the absolute derivative along the curve. For a specific timelike curve with tangent vector $U^{\mu}$, it can be expressed in terms of the covariant derivative  $\nabla_{\mu}$  as $\bd/\D \tau=U^{\mu}\nabla_{\mu}$ along the curve  (also valid in the region of interest, if $\tau$ can be extended to the entire region of interest\cite{liang2023differential}). The set of vectors $\{U^{\mu},\hat{A}^{\mu},\hat{B}^{\mu},\hat{C}^{\mu}\}$ at each specific value of $\tau$ forms a basis for the tangent space at a point in spacetime, and the normalization conditions of them are  
\begin{align}\label{normal}
	U^{\mu}U_{\mu}=-1,\qquad \hat{A}^{\mu}\hat{A}_{\mu}=\hat{B}^{\mu}\hat{B}_{\mu}=\hat{C}^{\mu}\hat{C}_{\mu}=1,
\end{align}
which is the direct consequence of the signature of spacetime. $\alpha, \delta, \eta$ are the parameters that characterize how the curve curves and twists in spacetime. The generalized Frenet-Serret formulas \eqref{FS-formula-GR} provide a basic construction of the local frame associated with a single observer. Spacetime imposes few constraints on a single observer, except when the observer approaches a boundary or singularity of spacetime.

However, in relativity, the principle that observations are inherently local implies that a single isolated observer is of limited utility, except when strictly comoving with the system under study. Typically, a physical process involves a finite region of spacetime, which necessitates the presence of a family of observers in general. An observer can be represented as  a timelike worldline equipped with a comoving local frame. A family of such observers naturally gives rise to a local frame field, and this frame field defines the reference frame.

With a family of observers, the issue of synchronization arises automatically. A physically reasonable condition minimally requires that a family of observers can synchronize their clocks. This requirement is closely related to the Frobenius theorem. Consider a family of observers whose worldlines cover the region of spacetime under investigation. The tangent vectors to these worldlines form a vector field, denoted as $Z^{\mu}\partial_{\mu}$, where $\partial_{\mu}$ represents the natural coordinate basis vectors. For the purpose of this discussion, the family of these observers will be referred to as $\mathscr{O}(Z)$. According to the Frobenius theorem, if the vector field $Z^{\mu}\partial_{\mu}$ satisfies
\begin{align}
	Z_{[\mu}\nabla_{\nu}Z_{\rho]}=0,
\end{align}
then $Z^{\mu}\partial_{\mu}$ must be orthogonal to a family of hypersurfaces in spacetime. (If one prefers to represent the observers using the one-form notation $\mathbf{Z}=Z_{\mu}\D x^{\mu}$, the condition of Frobenius theorem can be expressed in a more familiar form $\mathbf{Z}\wedge\D\mathbf{Z}=0$.) The hypersurfaces orthogonal to $Z^{\mu}\partial_{\mu}$ define the notion of simultaneity for the observers $\mathscr{O}(Z)$: events lying on the same hypersurface are simultaneous for this family. This provides a well-defined time evolution for physical systems observed by $\mathscr{O}(Z)$. Such a frame is called locally synchronizable, or hypersurface orthogonal \cite{sachs1977general}.

Intuitively, it may seem strange to impose conditions on the observers, but this is a necessary demand dictated by the principles of relativity theory. Without such constraints, obtaining a coherent description within a finite region would be challenging. Various additional requirements could be imposed to make the observers resemble those of the Newtonian mechanics, but such requirements are often too restrictive to be realizable in an arbitrary spacetime. So in later discussions, additional constraints are introduced only when necessary.

\subsection{Projection operator: Definition and Properties}
\label{sec-proj}

Having introduced hypersurface orthogonal observers, we now need to describe how physical information is represented from their perspectives. Such observers provide a natural foliation of spacetime, allowing us to project physical quantities onto the spatial hypersurfaces orthogonal to them. This is achieved by introducing projection operators.  

The general definitions of projection operators in spacetime are
\begin{align}\label{def-project}
	P_{\mu\nu}(n)&=\epsilon n_{\mu}n_{\nu},\\
	\Delta_{\mu\nu}(n)&=g_{\mu\nu}-\epsilon n_{\mu}n_{\nu},
\end{align}
where $n^{\mu}$ is a unit spacelike or timelike vector, and $\epsilon=g_{\mu\nu}n^{\mu}n^{\nu}$. One can easily verify that
\begin{align}
	P^{\mu}{}_{\nu}(n)n^{\nu}=n^{\mu},\qquad \Delta^{\mu}{}_{\nu}(n)n^{\nu}=0,
\end{align}
which indicates that $P^{\mu}{}_{\nu}(n)$ extracts the information of a geometric object along the vector $n^{\mu}$, while $\Delta^{\mu}{}_{\nu}(n)$ extracts the component orthogonal to $n^{\mu}$. As projection operators, they must also satisfy the property of idempotence,
\begin{align}
	P^{\mu}{}_{\rho}(n)P^{\rho}{}_{\nu}(n)=P^{\mu}{}_{\nu}(n),\qquad\Delta^{\mu}{}_{\rho}(n)\Delta^{\rho}{}_{\nu}(n)=\Delta^{\mu}{}_{\nu}(n).
\end{align} 

The vector field $Z^{\mu}\partial_{\mu}$ associated with observers $\mathscr{O}(Z)$ can be employed as the variable within the projection operator. The region of spacetime can be decomposed into a collection of spacelike hypersurfaces, with their sequence determined by the parameter of the integral curves of $Z^{\mu}\partial_{\mu}$. Minkowski provided  a more illustrative interpretation: `The three-dimensional geometry becomes a chapter of four-dimensional physics $\cdots$' Any type of tensor field, except for the scalar field, can be decomposed using the projection operator 
\begin{align}\label{decompose-T}
	T^{\mu_{1}\mu_{2}...\mu_{m}}{}_{\nu_{1}\nu_{2}...\nu_{n}}=\[P^{\mu_{1}}{}_{\rho}(Z)+\Delta^{\mu_{1}}{}_{\rho}(Z)\]\[P^{\sigma}{}_{\nu_{1}}(Z)+\Delta^{\sigma}{}_{\nu_{1}}(Z)\]T^{\rho\mu_{2}...\mu_{m}}{}_{\sigma\nu_{2}...\nu_{n}}.
\end{align}
These quantities obtained through such decomposition are closely related to the observable data as detected by the observers $\mathscr{O}(Z)$.

Indeed, when $\mathscr{O}(Z)$ represents the hypersurface orthogonal observers, the projection operator $\Delta_{\mu\nu}(Z)$ serves as the induced metric on the hypersurface orthogonal to $Z^{\mu}$. Consequently, the resulting notion of spatial distance becomes observer-dependent. Notice that the definition of a projection operator do not impose such constraints on its arguments. Therefore, defining a projection operator for a vector field $\tilde{Z}^{\mu}$ that is not hypersurface orthogonal is still permissible. However, in such a case, the geometric interpretation of the operation becomes ambiguous.

Finally, this section concludes by addressing a potential reason for the confusion between observers and coordinate systems. The variation of $\Delta_{\mu\nu}(Z)$ along the evolution of observers $\mathscr{O}(Z)$ can be depicted by the Lie derivative,  
\begin{align}
	\pounds_{Z}\Delta_{\mu\nu}(Z)&=Z^{\rho}\nabla_{\rho}(Z_{\mu}Z_{\nu})+\nabla_{\mu}Z_{\nu}+\nabla_{\nu}Z_{\mu}\notag\\
	&=2Z^{\rho}Z_{\mu}\nabla_{(\rho}Z_{\nu)}+2Z^{\rho}Z_{\nu}\nabla_{(\rho}Z_{\mu)}+2\nabla_{(\mu}Z_{\nu)}.
\end{align}
If $Z^{\mu}$ is Killing vector, the right-hand side of this equation vanishes, implying that the isometry of the hypersurface along the observers aligns with the isometry of spacetime. In this sense, the isometry of spacetime can be seen as inducing the isometry of hypersurface. A more interesting case arises when $Z^{\mu}$ satisfies the geodesic equation $Z^{\mu}\nabla_{\mu}Z_{\nu}=0$, but is not a Killing vector field. Observers whose proper velocity is given by such $Z^{\mu}$ are referred to as geodesic observers. If $\mathscr{O}(Z)$ is geodesic observers, then the first term on the right-hand side of the first line vanishes. The degree to which the observers align with the spacetime symmetry dictates how the induced metric varies from their perspectives.

A static observer in Cartesian coordinate in Minkowski spacetime is precisely a geodesic observer, with the tangent vector fields $Z^{\mu}$ a Killing vector field. It is therefore not surprising that, in the early stages of research on relativity, many concepts became intertwined.

\subsection{Uniformly Accelerated Observers}
\label{sec2.3}

When it comes to uniformly accelerated reference frame, a more common way to start the discussion is to write the metric as
\begin{align}\label{rindler}
	\D s^{2}=-X^{2}\D T^{2}+\D X^{2}+\D Y^{2}+\D Z^{2},
\end{align}
or in  some other equivalent forms\footnote{Many equivalent forms of this metric can be found in the item `Rindler coordinate' of Wikipedia.}. At first glance, it may seem to be unrelated to the discussion of the observers above, but this is not the case. It directly associates the natural coordinate basis with the observers, making this connection less apparent. This approach will later be referred to as the coordinate representation. This subsection uses this simple example to illustrate how to associate a coordinate system with a reference frame introduced above. The discussion begins with the simplest case of Minkowski spacetime.

In Minkowski spacetime with metric $\D s^{2}=\eta_{\mu\nu}\D x^{\mu}\D x^{\nu}$, where $\eta_{\mu\nu}=\text{diag\ }(-1,1,1,1)$, $x^{\mu}=(t,x,y,z)$, the Frenet-Serret formula reduce to
\begin{align}
	\dfrac{\D}{\D \tau}\begin{pmatrix}
		U^{\mu}\\
		\hat{A}^{\mu}\\
		\hat{B}^{\mu}\\
		\hat{C}^{\mu}
	\end{pmatrix}=
	\begin{pmatrix}
		0&\alpha &0&0\\
		\alpha &0&\delta &0\\
		0&-\delta &0&\eta\\
		0&0&-\eta &0
	\end{pmatrix}
		\begin{pmatrix}
		U^{\mu}\\
		\hat{A}^{\mu}\\
		\hat{B}^{\mu}\\
		\hat{C}^{\mu}
	\end{pmatrix},
\end{align}
where $\D/\D \tau$ is the absolute derivative along a certain worldline in Minkowski spacetime, and the normalized conditions in Eq.~\eqref{normal} are satisfied. The uniform acceleration implies the requirement
\begin{align}
	g_{\mu\nu}A^{\mu}A^{\nu}=g_{\mu\nu}\dfrac{\D U^{\mu}}{\D \tau}\dfrac{\D U^{\nu}}{\D \tau}=\alpha^{2}g_{\mu\nu}\hat{A}^{\mu}\hat{A}^{\nu}=\alpha^{2}=\text{const.}
\end{align}
Then by solving the Frenet-Serret formula, the resulting curve describes uniformly accelerated motion. However, for simplicity, additional conditions are often imposed. For example, if one is interested only in motion within a plane, then $\delta=0,\eta=0$ (or $\hat{B}^{\mu}$ and $\hat{C}^{\mu}$ are constant vectors along the worldline) need to be applied. Although these additional conditions are not necessary for uniformly accelerated motion, the extra constraints enforce $U^{\mu}$ and $\hat{A}^{\mu}$ remain in one plane\cite{rindler1960hyperbolic}, significantly simplifying the subsequent discussion. And then this equation can be written in a more concise form
\begin{align}
	\dfrac{\D^{2} U^{\mu}}{\D \tau^{2}}=\alpha^{2}U^{\mu}, 
\end{align}
together with the normalized condition $U_{\mu}U^{\mu}=-1$, an obvious solution can be obtained as 
\begin{align}
	U^{\mu}=T^{\mu}\cosh\alpha\tau +S^{\mu}\sinh\alpha\tau ,
\end{align}
where $T^{\mu}$ is a normalized timelike vector, $S^{\mu}$ is a normalized spacelike vector, and both of them are independent of the proper time $\tau$. Due to the introduced extra constraints above, if the plane of $U^{\mu}$ and $\hat{A}^{\mu}$ is chosen to be the $t-x$ plane, then the directions of $y$ and $z$ become redundant, essentially reducing the problem to a two-dimensional one. In 1+1 dimensional case, the simplest choices of $T^{\mu}$ and $S^{\mu}$ are
\begin{align}
	T^{\mu}=(1,0),\qquad S^{\mu}=(0,1).
\end{align}
Then the integral curve of $U^{\mu}$ can be obtained,
\begin{align}
	x^{\mu}=\dfrac{1}{\alpha}(\sinh\alpha\tau+\alpha C_{1},\cosh\alpha\tau+\alpha C_{2}),
\end{align}
where the $C_{1}$ and $C_{2}$ are integration constant determined by the initial conditions. Or in a more familiar form
\begin{align}
	-(t-C_{1})^{2}+(x-C_{2})^{2}=\dfrac{1}{\alpha^{2}},
\end{align}
which is the equation of a hyperbolic curve, obviously. Such a  simple discussion shows the profound difference between relativity and Newtonian mechanics, as in Newtonian mechanics, the curve representing uniformly accelerated motion in the $t-x$ plane is a parabola. $C_{1}$ and $C_{2}$ do not affect the shape of the curve, and $\alpha$ controls the curvature of the curve. A similar discussion can be carried out in the context of GR \cite{marder1957uniform,rindler1960hyperbolic}. 

Varying the parameter $\alpha$ generates a family of curves that covers a portion of Minkowski spacetime. It is then natural to take $\alpha=\text{const.}$ as the new coordinate lines $X$. Set $C_{1}=0$ and $C_{2}=0$, the equation of curves changes to
\begin{align}
	-t^{2}+x^{2}=X^{2},
\end{align} 
and then introduce another coordinate $T$ to describe the evolution along the curve, we have
\begin{align}\label{rindler-coord-trans}
	\left\{\begin{aligned}
		t&=X\sinh T,\\
		x&=X\cosh T.
	\end{aligned}\right.
\end{align}
This is actually a coordinate transformation for uniform acceleration. It is easy to obtain the metric of the portion of Minkowski space in this coordinate system
\begin{align}
	\D s^{2}=-X^{2}\D T^{2}+\D X^{2}, \qquad(-t^{2}+x^{2}>0).
\end{align}
Fulling  called the region of Minkowski spacetime covered by this coordinate system Rindler space  \cite{fulling1973nonuniqueness}. Every line satisfying $X=\text{const.}$ actually represents a uniformly accelerated observer with a magnitude of acceleration $\alpha=1/X$.

Introducing the new coordinate $T$ in the transformation Eq.~\eqref{rindler-coord-trans} implicitly assumes that the family of accelerated observers can be synchronized. This assumption is justified by verifying that the observers associated with the Rindler coordinates satisfy the hypersurface-orthogonality condition. The worldlines of static observers in Rindler coordinate are the integral curve of $\partial_{T}$. The normalized condition gives the reasonable vector fields of the family of static observers
\begin{align}
	Z^{\mu}\partial_{\mu}=\dfrac{1}{\sqrt{-t^{2}+x^{2}}}\(x\partial_{t}+t\partial_{x}\)=\dfrac{1}{X}\partial_{T}.
\end{align}
It can be readily verified that the Frobenius condition $Z_{[\mu}\nabla_{\nu}Z_{\rho]}=0$ is satisfied. For a single observer, replacing $\alpha \tau$ by the coordinate $T$ is just a linear reparameterization. However, performing such a transformation to a family of curves does not necessarily yield a family of synchronizable observers. An intuitive example will be provided later in the discussion of rotating observers.

In above discussion, the key idea is using the worldline of observers as timelike coordinate curves. To connect a well-defined local reference frame with the coordinate system, it may need to impose some constraints on the family of observers. Although it may sound wired, the existence of a well-defined observer depends on the properties of spacetime itself.

This discussion also highlights a fundamental distinction between the description of observers and that of particle motion. For observers, the worldlines are typically prescribed in advance, and the focus lies on their intrinsic geometric properties: whether they are static or stationary, inertial or accelerating, and whether they admit synchronization. In contrast, the study of particle motion is concerned with determining how particles move under given interactions. This involves solving equations of motion to obtain their worldlines. Once determined, these trajectories can, of course, be reinterpreted as the worldlines of hypothetical observers. However, note that observers are defined kinematically: the cause of their acceleration is not addressed, and no dynamical equations govern their motion.

In closing, while the coordinate representation provides a convenient means of connecting observers to the description of spacetime, it has an evident shortcoming. It often obscures the fact that experimental data are inherently associated with observers, rather than with the coordinate system itself. Distinguishing the coordinate system from the description of observers is a concept of fundamental importance and will be discussed further in the context of its historical development.

\section{Revisiting relativistic mechanics}
\label{sec3}

\subsection{The transformation of velocity and acceleration}

The first problem addressed by relativity is how physical quantities transform between different reference frames. In Einstein's first paper on relativity, he not only discussed length contraction and time dilation but also addressed the transformation of velocity and acceleration in the final sections\cite{einstein1905elektrodynamik}. The original discussions about the transformation all used the language of infinitesimals(cf. \cite{einstein1905elektrodynamik,planck1906prinzip,einstein1907relativity}), until the language of 4-dimensional geometry was introduced by Minkowski\cite{minkowski1909}. Under the framework of group theory, the invariance under coordinate transformations reflects the geometric essence of relativity. What this framework does not capture, however, is the independence of observers from the coordinates themselves. Therefore, in this section, we reformulate the transformation between different observers\cite{de2010classical}.

Consider a particle moving with velocity $U^{\mu}\partial_{\mu}$. Two different family of observers $\mathscr{O}(Y)$ and $\mathscr{O}(Z)$ described by vector fields $Y^{\mu}$ and $Z^{\mu}$. According to the projection operator introduced in  last section, the proper velocity of the particle can be decomposed into
\begin{align}\label{decompose-U}
	U^{\mu}&=P^{\mu}{}_{\rho}(Y)U^{\rho}+\Delta^{\mu}{}_{\rho}(Y)U^{\rho}=\gamma_{uy}[Y^{\mu}+u^{\mu}(Y)]\notag\\
	&=P^{\mu}{}_{\rho}(Z)U^{\rho}+\Delta^{\mu}{}_{\rho}(Z)U^{\rho}=\gamma_{uz}[Z^{\mu}+u^{\mu}(Z)],
\end{align}
where $\gamma_{uy}=-U_{\rho}Y^{\rho},\gamma_{uz}=-U_{\rho}Z^{\rho}$, and
\begin{align}
	u^{\mu}(Y)=\gamma_{uy}^{-1}\Delta^{\mu}{}_{\rho}(Y)U^{\rho},\qquad u^{\mu}(Z)=\gamma_{uz}^{-1}\Delta^{\mu}{}_{\rho}(Z)U^{\rho}.
\end{align}
According to the normalization condition of proper velocity, 
\begin{align}
	-1=U^{\mu}U_{\mu}=\gamma_{uy}^{2}\[Y^{\mu}+u^{\mu}(Y)\]\[Y_{\mu}+u_{\mu}(Y)\]=\gamma_{uy}^{2}\[u^{\mu}(Y)u_{\mu}(Y)-1\],
\end{align}
from which one can deduce
\begin{align}
	\gamma_{uy}=\pm\dfrac{1}{\sqrt{1-u^{2}(Y)}}.
\end{align}
And to ensure that the increment direction of the proper time for both particles and observers aligns, the negative sign in the equation is typically omitted. The ratio between the proper time of observers and particles is related by the local Lorentz factor
\begin{align}\label{dt-dtau}
	\D t_{Y}=\dfrac{\partial t_{Y}}{\partial x^{\mu}}\D x^{\mu}=-\lambda_{Y} Y_{\mu}\D x^{\mu}=-\lambda_{Y} Y_{\mu}U^{\mu}\D \tau=\lambda_{Y}\gamma_{uy}\D \tau,
\end{align}
where $\lambda_{Y}=|\nabla t_{Y}|$ is the normalization factor ensuring that $Y^{\mu}\partial_{\mu}$ is a unit timelike vector. This factor quantifies the relation between coordinate time and the proper time of observers, once again highlighting the distinction between reference frames and coordinate systems.

The necessity of considering transformations arises from the fact that a particle's behavior is observer-dependent. Similarly, the behavior of a given family of observers can itself be analyzed from the perspective of another family via their respective projection operators. For example, the velocity field of $\mathscr{O}(Y)$ can be decomposed as
\begin{align}\label{y-z}
	Y^{\mu}=\gamma_{yz}[Z^{\mu}+y^{\mu}(Z)],
\end{align}
where 
\begin{align*}
	\gamma_{yz}=-Y_{\rho}Z^{\rho},\qquad y^{\mu}(Z)=\gamma_{yz}^{-1}\Delta^{\mu}{}_{\rho}(Z)Y^{\rho}.
\end{align*}
And then the geometric object observed by observers $\mathscr{O}(Y)$ can be expressed by the observations of $\mathscr{O}(Z)$. For instance, the velocity of a particle observed by $\mathscr{O}(Z)$, when transformed to the reference frame of observers $\mathscr{O}(Y)$(both expressed in a fixed coordinate system), is given by
\begin{align}\label{uy-uz}
	u^{\mu}(Y)&=\gamma_{uy}^{-1}\gamma_{uz}\Delta^{\mu}{}_{\nu}(Y)\[u^{\nu}(Z)+Z^{\nu}\]\notag\\
	&=\[\gamma_{uy}^{-1}\gamma_{uz}u^{\mu}(Z)-\gamma_{yz}y^{\mu}(Z)\]+\(\gamma_{uy}^{-1}\gamma_{uz}-\gamma_{yz}\)Z^{\mu},
\end{align}
where in the first line $U^{\mu}$ is decomposed by $Z^{\mu}$, and in the second line the operator $\Delta^{\mu}{}_{\nu}(Y)$ is further decomposed in terms of $Z^{\mu}$ to obtain an explicit expression.
                         
Some careful reader may notice that this equation differs from the usual form of transformation in SR, even when expressed in a concrete coordinate system. The reason is that the observers are usually binding with coordinate systems in SR. To recover the conventional results, an extra coordinate transformation matrix is needed,
\begin{align}
	u^{\mu}(Y)\partial_{\mu}=\Lambda^{\mu}{}_{\nu}\tilde{u}^{\nu}(Y)(\Lambda^{-1})^{\rho}{}_{\mu}\tilde{\partial}_{\rho},
\end{align}
where $\Lambda^{\mu}{}_{\nu}$ transforms vector components from the coordinate system adapted to $\mathscr{O}(Y)$ to that adapted to $\mathscr{O}(Z)$, and $\tilde{u}^{\mu}$ denotes the velocity components as they commonly appear in standard textbooks on special relativity. However, it is important to note that $u^{\mu}(Y)\partial_{\mu}$ represents the true geometry object, while both $u^{\mu}$ and $\tilde{u}^{\mu}$ are merely the components in different coordinate systems.

The proper acceleration of the particle is defined as
\begin{align}
	A^{\mu}=U^{\nu}\nabla_{\nu}U^{\mu},
\end{align}
which is again an observer independent object. The observer-dependence arises from the decomposition within the reference frame defined by observers, and it can be written as 
\begin{align}
	A^{\mu}&=A^{\nu}Y_{\nu}Y^{\mu}+\Delta^{\mu}{}_{\nu}(Y)A^{\nu}\notag\\
    &=A^{\nu}Z_{\nu}Z^{\mu}+\Delta^{\mu}{}_{\nu}(Z)A^{\nu}.
\end{align}

However, the decomposition of acceleration is more subtle than velocity, as it involves the second-order derivative with respect to proper time. The question here is how to define the observed acceleration. If the acceleration is defined as the rate of change of proper velocity with respect to the increasing of proper time of the observers $\mathscr{O}(Y)$, then it can be expressed as
\begin{align}
	\mathcal{A}^{\mu}(Y)=\dfrac{\bd}{\D t_{Y}}U^{\mu}=\dfrac{\D }{\D t_{Y}}U^{\mu}+\varGamma^{\mu}_{\rho\sigma}Y^{\rho}U^{\sigma}=Y^{\nu}\nabla_{\nu}U^{\mu},
\end{align}
where $\bd/\D t_{Y}$ are the directional derivative along the worldline of the observers $\mathscr{O}(Y)$. $\mathcal{A}^{\mu}(Z)$ can be defined in the same way. In this way, the relationship between the proper acceleration $A^{\mu}$ and the observed acceleration $\mathcal{A}^{\mu}$ is 
\begin{align}
	A^{\mu}&=\gamma_{uy}[Y^{\nu}+u^{\nu}(Y)]\nabla_{\nu}U^{\mu}=\gamma_{uy}\mathcal{A}^{\mu}(Y)+\gamma_{uy}u^{\nu}(Y)\nabla_{\nu}U^{\mu},\notag\\
    &=\gamma_{uz}[Z^{\nu}+u^{\nu}(Z)]\nabla_{\nu}U^{\mu}=\gamma_{uz}\mathcal{A}^{\mu}(Z)+\gamma_{uz}u^{\nu}(Z)\nabla_{\nu}U^{\mu},
\end{align}
The transformation between two observers is
\begin{align}
	\mathcal{A}^{\mu}(Y)&=\gamma_{yz}[Z^{\nu}+y^{\nu}(Z)]\nabla_{\nu}U^{\mu}\notag\\
	&=\gamma_{yz}\mathcal{A}^{\mu}(Z)+\gamma_{yz}y^{\nu}(Z)\nabla_{\nu}U^{\mu}.
\end{align}
Such a definition simplifies the discussion of transformations, as it relies directly on the concept of decomposition.

At the early stage, however, the transformation of acceleration is based on its definition within Newtonian framework, which, in present language, should be expressed as 
\begin{align}
	a^{\mu}(Y)=\dfrac{\bd}{\D t_{Y}}u^{\mu}(Y)=Y^{\nu}\nabla_{\nu}u^{\mu}(Y),
\end{align}
and $a^{\mu}(Z)$ follows the same form. Now combine with eq.~\eqref{y-z} and eq.~\eqref{uy-uz}, the transformation between $a^{\mu}(Y)$ and $a^{\mu}(Z)$ is obvious. Due to the notion of relativity and Newtonian framework are conflated and discussed together here, the resulting analysis is inevitably ugly. In this regard, the older version is not significantly better\cite{pauli1921theory}.

The result is not explicitly stated here, as it is not necessary; however, it is worth noting why the analysis of acceleration is distinct in the context of kinematic discussions. The result of decomposition in acceleration contains the derivative of the proper velocity of observers, for example, the terms like $\nabla_{\mu}Y^{\nu}$ for observers $\mathscr{O}(Y)$. Undoubtedly, these terms implicitly foreshadow the equivalence principle.

The discussion in this section may not fully reveal the convenience of the projection operator method within SR. This limitation is caused by the degenerate nature of SR. In GR, coordinates are, in principle, not intrinsically tied to reference frames, compelling us to address issues in this way. Moreover, the projection operator method offers significant interpretative advantages: transformations therefore merely involve vector decomposition, which, while not necessarily simplifying calculations, eliminates the need for case-by-case analysis in conceptual understanding. Perhaps Einstein's words provide the best defense for the discussion in this section: `I agree that my mathematical tool is more complicated than yours, but my physical assumptions are simpler and more natural.' Originally, this statement was used in defense of general relativity\cite{einstein1966evolution}.

\subsection{Motion of a Single Particle}

This subsection focuses on the aspects of particle motion that pertain to observers. The main goal is to elucidate why the formalism developed in the early stages of relativity was derivable even before the notion of an observer was rigorously defined, and to see how these results paved the way for the geometric formulation of general relativity. The formulations provided by analytical mechanics have significantly advanced the development of classical mechanics, both conceptually and operationally. In the development of relativity, it also plays a significant role.

It can be expected that the description of moving particles should be governed by equations analogous to Newton's second law, but now in SR, the governing group is the Poincaré group rather than the Galileo group. The description should be coordinate-free manner, and consequently, the natural starting point is the generalization of the principle of stationary action. It is well-known in classical mechanics, the action of a conservative system is the integral of Poincaré 1-form $p_{i}\D x^{i}$ on cotangent bundle\cite{arnold1978mathematical}. Therefore, it is reasonable to expect that the generalization of Poincaré 1-form\footnote{Numerous textbooks on relativity prefer to start with $S=\int\D s$, {\it i.e.} taking the length of worldline as the action. While this approach yields results equivalent to those obtained from the Poincaré 1-form, the method adopted here offers certain advantages: (1) it provides a natural transition from the Newtonian framework to relativistic mechanics; (2) it avoids the complications associated with the square root, which typically requires lengthy explanations to address.} should take the form $p_{\mu}\D x^{\mu}$. The action immediately follows the generalization,
\begin{align}\label{action}
	S=&\int p_{\mu}\D x^{\mu}=\int \mathscr{L}_{\tau}\D \tau=\int \mathscr{L}_{t}\D t,
\end{align}
where the second equality is a naive generalization of Lagrangian formalism, and the third equality is the one used in the early stage. It is a common viewpoint that formulations expressed in terms of a splitting of spacetime breaks the symmetry of relativity, but this is not the fact. Once the time $t$ in eq.~\eqref{action} is recognized as the coordinate time, significant effort is required to demonstrate that the theory remains coordinate-free. But if the time $t$ is understood as the proper time of observers $\mathscr{O}(Z)$(actually, it is the proper time of a single observer, and the proper time for the entire reference frame is obtained by extending $t$ over the family of hypersurface orthogonal observers), the covariance required by relativity is not affected at all. Of course, these different understandings lead to differences in expressions of the formalism. The following part will show how to apply the variational principle under the introduction of observers in SR. 

Firstly, it is necessary to clarify the basic variables of Lagrangian. Lagrangian is thought to be a function on tangent bundle of Euclidean space in classical mechanics, hence its generalization is naturally a function on tangent bundle $T\mathcal{M}$ of spacetime $\mathcal{M}$. When take the proper time of the particle as the evolution parameter, the arguments of Lagrangian can be chosen as $\{x^{\mu},\dot{x}_{\tau}^{\mu}\}$, while take the proper time of the observers $\mathscr{O}(Z)$ as the evolution parameter, the arguments can be chosen as $\{x^{\mu},\dot{x}_{t}^{\mu}\}$, where 
\begin{align}
	\dot{x}_{\tau}^{\mu}\equiv\dfrac{\D x^{\mu}}{\D \tau}\equiv U^{\mu},\qquad \dot{x}_{t}^{\mu}\equiv\dfrac{\D x^{\mu}}{\D t}.
\end{align}
Notice that $\dot{x}_{t}^{\mu}(t)$ is still a vector in the tangent space of spacetime. According to eq.~\eqref{dt-dtau}, we have 
\begin{align}\label{U-xt}
	U^{\mu}\equiv\dfrac{\D x^{\mu}}{\D \tau}=\dfrac{\D x^{\mu}}{\D t}\dfrac{\D t}{\D \tau}=\lambda\gamma_{uz}\dot{x}_{t}^{\mu},
\end{align}
which prove the assertion, and we can see $\dot{x}_{t}^{\mu}(t)$ is just a rescale of $\dot{x}_{\tau}^{\mu}(\tau)$. Together with $U^{\mu}=\gamma_{uz}[Z^{\mu}+u^{\mu}(Z)]$, we obtain a useful equation in the following deduction,
\begin{align}\label{xt-Z}
	\lambda\dot{x}_{t}^{\mu}=Z^{\mu}+u^{\mu}(Z).
\end{align}

Stationary action principle gives the Euler-Lagrange equation,
\begin{align}
	\delta S&=\int\delta\mathscr{L}_{t}(x^{\mu},\dot{x}_{t}^{\mu})\D t\notag\\
    &=\int\(\delta x^{\mu}\partial_{\mu}\mathscr{L}_{t}+\delta\dot{x}^{\mu}\dfrac{\partial\mathscr{L}_{t}}{\partial\dot{x}^{\mu}}\)\D t\notag\\
	&=\left.\delta x^{\mu}\dfrac{\partial\mathscr{L}_{t}}{\partial \dot{x}^{\mu}}\right|_{\text{boundary}}-\int\[\dfrac{\D}{\D t}\(\dfrac{\partial\mathscr{L}_{t}}{\partial \dot{x}^{\mu}}\)-\partial_{\mu}\mathscr{L}_{t}\]\delta x^{\mu}\D t=0,
\end{align}
where $\mathscr{L}_{t}(x^{\mu},\dot{x}_{t}^{\mu})$ is a function on tangent bundle $T\mathcal{M}$ of spacetime $\mathcal{M}$. Following the standard procedure, the value of coordinate at boundary is fixed so that the general form of Euler-Lagrange equation could be written as
\begin{align}\label{EL-ab}
	\dfrac{\D}{\D t}\dfrac{\partial \mathscr{L}}{\partial \dot{x}_{t}^{\mu}}-\dfrac{\partial\mathscr{L}}{\partial x^{\mu}}=0.
\end{align}
Some additional remarks are in order. Note that variational operator $\delta$ and the differentiation along a curve $\D/\D t$ are different operations, and they do not even act on the object in the same space. For synchronous variations, the commutation of $\D/\D t$ and $\delta$ follows directly from their definitions. At the same time, one should avoid confusing $\D/\D t$ with the $\bd/\D t$ introduced earlier, which refers to the absolute derivative used for parallel transportation of vectors on a manifold. Here, $\D/\D t$ is only a measurment of the variation of a quantity along the curve in observers' time.\footnote{In dealing with variational problems, it is never necessary to introduce  $\bd/\D t$. The affine connection terms naturally emerge from applying $\D/\D t$ to the metric. These terms encapsulate the geometric structure of spacetime and ultimately ensure that the resulting equations are covariant. For the proof of the covariance of the equation, refer to Appendix \ref{appA}.}

Let us shift focus from the abstract discussion and turn attention to the derivation of the equations of motion for a charged particle in an electromagnetic field. When a charged particle is moving in an electromagnetic field $F_{\mu\nu}\equiv\partial_{\mu} A_{\nu}-\partial_{\nu}A_{\mu}$, the kinematic momentum in Poincaré 1-form should be replaced by the canonical momentum
\begin{align}
	P_{\mu}\equiv p_{\mu}+qA_{\mu}.
\end{align}
Then the Lagrangian can be written down as follows
\begin{align}
	\mathscr{L}_{t}=p_{\mu}\dfrac{\D x^{\mu}}{\D t}+qA_{\mu}\dfrac{\D x^{\mu}}{\D t}.
\end{align}
Together with the definition of kinematic momentum $p^{\mu}=mU^{\mu}$ and eq.~\eqref{U-xt}, the Lagrangian reduced to 
\begin{align}
	\mathscr{L}_{t}&=m\lambda\gamma_{uz}\eta_{\mu\nu}\dot{x}_{t}^{\mu}\dot{x}_{t}^{\nu}+qA_{\mu}\dot{x}_{t}^{\mu}\notag\\
	&=\dfrac{m\gamma_{uz}}{\lambda}[Z_{\mu}+u_{\mu}(Z)][Z^{\mu}+u^{\mu}(Z)]+qA_{\mu}\dot{x}_{t}^{\mu}\notag\\
	&=-\dfrac{m}{\lambda\gamma_{uz}}+qA_{\mu}\dot{x}_{t}^{\mu},
\end{align}
where eq.~\eqref{xt-Z} is substituted in the second line, and
\begin{align}
	\gamma_{uz}=\dfrac{1}{\sqrt{1-u^{2}(Z)}}=\dfrac{1}{\sqrt{1-\eta_{\mu\nu}\(\lambda\dot{x}_{t}^{\mu}-Z^{\mu}\)\(\lambda\dot{x}_{t}^{\nu}-Z^{\nu}\)}}.
\end{align}
With the explicit form of the Lagrangian, deriving the equations of motion becomes a straightforward procedural task. First, we have
\begin{subequations}
\begin{align}
	\dfrac{\partial\mathscr{L}_{t}}{\partial \dot{x}_{t}^{\mu}}&=\dfrac{mu_{\mu}(Z)}{\sqrt{1-u^{2}(Z)}}+qA_{\mu},\label{pL/dx}\\
	\dfrac{\partial\mathscr{L}_{t}}{\partial x^{\mu}}&=\dfrac{m}{\gamma_{uz}\lambda^{2}}\partial_{\mu}\lambda-\dfrac{m}{\lambda}\partial_{\mu}\gamma_{uz}^{-1}+q\dot{x}_{t}^{\nu}\partial_{\mu}A_{\nu},\label{pL/x}
\end{align}
\end{subequations}
where the second term on the right-hand side of eq.~\eqref{pL/x} can be further simplified,
\begin{align}\label{pg-1}
	\partial_{\mu}\gamma_{uz}^{-1}&=-\dfrac{1}{2\sqrt{1-u^{2}(Z)}}\partial_{\mu}\[\eta_{\rho\sigma}\(\lambda \dot{x}_{t}^{\rho}-Z^{\rho}\)\(\lambda \dot{x}_{t}^{\sigma}-Z^{\sigma}\)\]\notag\\
	&=-\gamma_{uz}u_{\rho}(Z)\(\dot{x}_{t}^{\rho}\partial_{\mu}\lambda-\partial_{\mu}Z^{\rho}\)\notag\\
	&=-\dfrac{\gamma_{uz}\partial_{\mu}\lambda}{\lambda}u_{\rho}(Z)[Z^{\rho}+u^{\rho}(Z)]-\gamma_{uz}Z_{\rho}\partial_{\mu}u^{\rho}(Z)\notag\\
	&=-\dfrac{\gamma_{uz}u^{2}}{\lambda}\partial_{\mu}\lambda-\gamma_{uz}Z_{\rho}\partial_{\mu}\(\lambda\dot{x}_{t}^{\rho}-Z^{\rho}\)\notag\\
	&=-\dfrac{\gamma_{uz}u^{2}}{\lambda}\partial_{\mu}\lambda-\gamma_{uz}\dot{x}_{t}^{\rho}Z_{\rho}\partial_{\mu}\lambda\notag\\
	&=-\dfrac{\gamma_{uz}u^{2}}{\lambda}\partial_{\mu}\lambda-\dfrac{\gamma_{uz}\partial_{\mu}\lambda}{\lambda}Z_{\rho}[Z^{\rho}+u^{\rho}(Z)]\notag\\
	&=\dfrac{\gamma_{uz}\partial_{\mu}\lambda}{\lambda}\[1-u^{2}(Z)\]=\dfrac{\partial_{\mu}\lambda}{\lambda\gamma_{uz}}.
\end{align}
Substituting eq.~\eqref{pg-1} into eq.~\eqref{pL/x}, a more concise form of eq.~\eqref{pL/x} is obtained
\begin{align}
	\dfrac{\partial\mathscr{L}_{t}}{\partial x^{\mu}}&=q\dot{x}_{t}^{\nu}\partial_{\mu}A_{\nu}.
\end{align}
The time derivative of eq.~\eqref{pL/dx} gives the first part of Euler-Lagrange equation,
\begin{align}
	\dfrac{\D }{\D t}\(\dfrac{\partial\mathscr{L}_{t}}{\partial \dot{x}^{\mu}}\)=\dfrac{\D }{\D t}\(\dfrac{mu_{\mu}(Z)}{\sqrt{1-u^{2}(Z)}}\)+q\dot{x}_{t}^{\nu}\partial_{\nu}A_{\mu}.
\end{align}
Finally, combine with eq.~\eqref{pL/x}, the equation of motion of a charged particle in electromagnetic fields is given
\begin{align}
	\dfrac{\D }{\D t}\(\dfrac{mu^{\mu}(Z)}{\sqrt{1-u^{2}(Z)}}\)=qF^{\mu}{}_{\nu}\dot{x}_{t}^{\nu},
\end{align}
which is exactly the coordinate-free form corresponding to the equations obtained by Einstein, Planck, and others in the early stages.

We would like to reiterate that the purpose of this section is not to discuss how to derive the equations of motion, but rather to explore why early physicists did not distinguish between the observers and the coordinates but still obtained a lot of useful results. Based on the derivation above, it can be concluded that the variational principle is independent of the choice of the observer. The transformation from coordinate time to the proper time of the observer also does not change the form of the procedure greatly, though the meaning of every step in the derivation is totally different. The validity of the variational principle itself is independent of observers and coordinate systems, and it inherently captures the geometric nature of motion. This enables the development of relativistic mechanics to proceed in a more concise and elegant manner, which in some sense paves the way for Einstein's subsequent work from relativistic mechanics to GR\cite{einstein1912theory}.

\subsection{Ehrenfest paradox}

To define rigidity within Newtonian mechanics, it is convenient to adopt the Lagrangian picture of continuum mechanics. Continuum mechanics concerns about the motion of a continuum in the absolute space, while material particles in the continuum construct another space which refers to matter space. The rigidity then can be regarded as a property imposed on the matter space. Let $h_{ij}$ denote the metric of the matter space, and $u^{i}$ represent the velocity field of the continuum. The condition of rigidity requires 
\begin{align}
	\mathcal{L}_{u}h_{ij}=u^{k}\partial_{k}h_{ij}+h_{kj}\partial_{i}u^{k}+h_{ik}\partial_{j}u^{k}=0.
\end{align}
Generally, the metric of the matter space is taken as the Euclidean metric $h_{ij}\equiv\eta_{ij}$, ensuring that the notion of distance therein is consistent with our Newtonian intuition. It is constant tensor in Cartesian coordinate system, so the rigidity condition reduces to the familiar form
\begin{align}
	\partial_{i}u_{j}+\partial_{j}u_{i}=0,
\end{align}
which directly expresses the fact that the distance between any two material points remains invariant during the motion.

The intuitive ideas about the generalization of rigidity is then\cite{rosen1947notes} 
\begin{align}\label{def-rigid}
	\mathcal{L}_{U}\Delta_{\mu\nu}(U)=U^{\rho}\nabla_{\rho}(U_{\mu}U_{\nu})+\nabla_{\mu}U_{\nu}+\nabla_{\nu}U_{\mu}=0,
\end{align}
where $U^{\mu}$ is the proper velocity field of the continuum, and $\Delta_{\mu\nu}(U)$ is the projection operator defined by the continuum. As previously mentioned, when $U^{\mu}\partial_{\mu}$ satisfies the Frobenius condition, $\Delta_{\mu\nu}(U)$ naturally becomes the metric of the spacelike hypersurfaces normal to $U^{\mu}\partial_{\mu}$. This definition aligns precisely with the concept of a rigid body in Newtonian framework. 

However, in the relativistic framework---especially in the early days when observers and coordinate systems were often conflated—--it sparked significant controversy. Here, we highlight two key points: 1. Since eq.~\eqref{def-rigid} is a tensor equation, the definition of rigidity is coordinate-free; 2. For the notion of relativistic rigidity to align with its Newtonian counterpart, eq.~\eqref{def-rigid} must implicitly assume the existence of a family of hypersurface orthogonal comoving observers $\mathscr{O}(Z)\equiv\mathscr{O}(U)$. Without such observers, a definition of rigidity is still possible, but its interpretation would depart significantly from the intuitive Newtonian understanding.

After Born proposed his definition of rigid body in relativity and provided the hyperbolic motion of electron as an example\cite{born1909theorie,born1910definition,born1910kinematik}, it quickly faced significant criticism\cite{ehrenfest1909gleichformige,ehrenfest1910herrn,ehrenfest1911herrn,laue1911diskussion,herglotz1910standpunkt,herglotz1911mechanik,planck1910gleichformige}. One of the most intuitive objections being Ehrenfest's example of a uniformly rotating disk in Minkowski spacetime. A disk is an object with cylindrical symmetry that occupies a finite region of spacetime. Given its symmetry, the cylindrical coordinate system is the most natural choice for its description. The metric of Minkowski spacetime in this coordinate system is given by
\begin{align}\label{Min-cy-metric}
	\D s^{2}=-\D t^{2}+\D r^{2}+r^{2}\D \varphi^{2}+\D z^{2}.
\end{align}
Ehrenfest discovered an intriguing fact: static observers in the coordinate system mentioned above will measure a different circumference for the disk compared to observers who are comoving with the uniformly rotating disk. It is precisely at this point that the observers' description was introduced. 

The worldline of a uniformly rotating observer is $x^{\mu}(t)=(t,r_{0},\varphi_{0}+\omega t,z_{0})$, where $r_{0}$, $\varphi_{0}$,and $z_{0}$ is the initial position of the observer and $\omega$ is the angular velocity. By varying the initial values of $x^{\mu}(t)$, we can obtain a family of observers, which we shall refer to as observers $\mathscr{O}(Z)$. The tangent vector field of observers $\mathscr{O}(Z)$ is
\begin{align}
	Z^{\mu}\partial_{\mu}=\dfrac{1}{\sqrt{1-\omega^{2}r^{2}}}\(\partial_{t}+\omega\partial_{\varphi}\).
\end{align} 
Following the same procedure used to introduce the coordinate representation for hyperbolic motion observers in Sec.~\ref{sec2.3}, the metric in this new coordinate system, constructed using the worldlines of uniformly rotating observers, can be written as
\begin{align}
	\D s^{2}=-\(1-\omega^{2} r_{0}^{2}\)\D t_{0}^{2}+\D r_{0}^{2}+r_{0}^{2}\D\varphi^{2}_{0}+2\omega r_{0}^{2}\D t_{0}\D\varphi_{0}+\D z_{0}^{2},
\end{align}
where the coordinate transformation
\begin{align*}
	t=t_{0},\qquad r=r_{0},\qquad \varphi=\varphi_{0}+\omega t_{0},\qquad z=z_{0},
\end{align*}
has been substituted into eq.~\eqref{Min-cy-metric}. In this new coordinate system, the tangent vector of observers $\mathscr{O}(Z)$ is $Z^{\mu}\partial_{\mu}=(1-\omega^{2}r_{0}^{2})^{-1/2}\partial_{t_{0}}$. It can be verified that, regardless of the coordinate system, this family of observers does not satisfy the Frobenius condition, {\it i.e.}
\begin{align*}
	Z_{[\mu}\nabla_{\nu}Z_{\rho]}\ne 0.
\end{align*}
Therefore, this family of observers is unsynchronizable and cannot constitute a valid reference frame. The reason Ehrenfest and his contemporaries arrived at the aforementioned result is that they overlooked the conditions related to synchronization. The metric of matter space is obtained by
\begin{align}
	\D s_{\text{disk}}^{2}&=\Delta_{\mu\nu}(U)\D x^{\mu}\D x^{\nu}=\Delta_{\mu\nu}(Z)\D x^{\mu}\D x^{\nu}\notag\\
	&=(\eta_{\mu\nu}+Z_{\mu}Z_{\nu})\D x^{\mu}\D x^{\nu}=\D r_{0}^{2}+\dfrac{r_{0}^{2}}{1-\omega^{2}r_{0}^{2}}\D \varphi_{0}^{2}+\D z_{0}^{2},
\end{align}
Compared to the spacelike part of metric \eqref{Min-cy-metric}, the circumference measured by the uniformly rotating observer can be determined as $(1-v^{2})^{-1/2}$  times the circumference measured by the static observer, where $v=\omega r_{0}$ and $r_{0}$ represents the radius of the disk as measured by the static observer.

Regarding the rigidity of a uniformly rotating disk, a direct calculation confirms that
\begin{align}\label{rigidity-def}
	\mathcal{L}_{U}\Delta_{\mu\nu}(U)=\mathcal{L}_{Z}\Delta_{\mu\nu}(Z)=0.
\end{align}
This covariant definition of rigidity was introduced by Rosen \cite{rosen1947notes}, who argued that`...it would be unsatisfactory if no rigid bodies or rigid-body motions existed in the theory'. As mentioned in Sec.~\ref{sec-proj}, when $Z^{\mu}\partial_{\mu}$ satisfies the Frobenius condition, the projection operator $\Delta_{\mu\nu}(Z)$ has a clear interpretation as the induced metric on the hypersurface. In the absence of the Frobenius condition, however, the same equations remain mathematically well-defined, but their physical interpretation becomes ambiguous.

This result has led to two sharply contrasting attitudes toward rigidity: one camp asserted that a relativistic treatment of rigid bodies is fundamentally impossible\cite{planck1910gleichformige}, while the other argued that, despite the potential issues of the definition of a relativistic rigid body, the concept itself was necessary, and at the very least, the discussion of rigid motion was permissible\cite{born1910kinematik,rosen1947notes}. At its core, the disagreement reflects the extent to which different physicists sought to preserve the Newtonian conception of rigidity within the relativistic framework. If one adopts \eqref{def-rigid} as the definition of rigidity, then for a uniformly rotating disk, the difficulty lies not in the failure of rigidity itself, but rather in the impossibility of synchronizing clocks among comoving observers. Some researchers who rejected the definition of rigidity took an alternative path, turning instead to the development of relativistic hydrodynamics and the theory of elasticity\cite{herglotz1911mechanik,laue1911dynamik}.

\section{History and Reflections}
\label{sec4}

\subsection{Historic timeline}

\begin{itemize}
	\item  1905: Einstein analyzed the transformations of length, time interval, velocity, and acceleration by introducing `rigid measuring rods' and binding them with Cartesian coordinates\cite{einstein1905elektrodynamik}.
	\item  1906: Planck quickly recognized that Einstein's work could be reformulated in an analytic mechanics form, including the equation of motion in both Lagrangian and Hamiltonian formulations are provided\cite{planck1906prinzip}. Planck repeatedly referred to and discussed it in several papers over the following years\cite{planck1908dynamik,planck1998eight}. 
	\item 1907:
	\begin{itemize}
		\item On March 19, Ehrenfest submitted a brief commentary to Annalen der Physik on Abraham's discussion concerning the forces acting on (rigid) non-spherical electrons in motion \cite{ehrenfest1907translation}. He posed the following question:`When in uniform translation,   according to Einstein  the electron undergoes the known Lorentz contraction. Now, is a uniform translation into every direction in a force-free manner for this electron possible, or not?'
		\item Einstein promptly responded in April \cite{einstein1907bemerkungen}. He clarified that relativity should be regarded as a principle theory, rather than a constructive one aimed at providing equations for specific physical processes. He also explicitly stated that both the dynamics and kinematics of rigid bodies remained undefined at the time.  
		\item A footnote in a  review article  in December  clarified his use of `rigid measuring rods' \cite{einstein1907onthe}, a concept that had frequently appeared in his earlier papers, though a rigorous definition had yet to be provided. In the same article, he devoted a section to discussing the challenges of reconciling the concept of rigid bodies with the principles of relativity.
		\item Planck's formulation in terms of analytical mechanics was taken up by Einstein \cite{einstein1907relativity}.
	\end{itemize}
	\item  1909: 
	\begin{itemize}
		\item Minkowski reformulated Einstein's special relativity using a geometric framework that revealed the underlying structure of the theory\cite{minkowski1909}. He introduced the concept of the {\it worldline} in his discussion. Minkowski also commented on the rigid body in Newtonian framework in his lecture, `On the other hand, the concept of a rigid body has meaning only in a mechanics with the group $G_{\infty}$.' Here, $G_{c}$ denotes the relativity group (the Poincaré group), while $G_{\infty}$ represents its Newtonian limit (the Galilean group).
		\item Born had conducted further discussions with the aid of the geometric language \cite{born1909theorie}. Born argued that the definition of a rigid body should also be connected to these transformations. 
		\item Ehrenfest quickly provided a brief discussion\cite{ehrenfest1909gleichformige}, which appeared to contradict Born's definition. 
	\end{itemize}
	\item   1910:
	\begin{itemize}
		\item Born's definition was rigorously checked by Herglotz\cite{herglotz1910standpunkt} and Noether\cite{noether1910kinematik}, but they conclude that such a definition is too restrictive to be useful. Born acknowledged the limitations of his initial definition, he remained reluctant to completely abandon his attempts to define rigidity in relativity\cite{born1910kinematik}.
		\item In a brief commentary \cite{planck1910gleichformige}, Planck remarked on the concerns raised by others, stating:`the attempt to make the abstraction of the rigid body (which is so important for ordinary mechanics) also useful for the theory of relativity, does not promise any real success.' His line of reasoning is essentially consistent with the ideas of Einstein in 1907\cite{einstein1907onthe}, and he also pointed out that the necessity of developing elasticity theory in relativity.
		\item Sommerfeld wrote a review paper introducing Minkowski's geometric language\cite{sommerfeld1910relativitatstheorie}.
	\end{itemize}
	\item   1911:
	\begin{itemize}
		\item The question of whether the differentials $\{\D t, \D x\}$ could be interpreted as representing a physically realized event for an object remained a point of contention \cite{varicak1911zum, einstein1911zum}. In these discussions, Varićak \cite{varicak1911zum} appears to have touched upon the fundamental distinction between coordinate systems and observers, though his views met with strong opposition from Einstein \cite{einstein1911zum}. Einstein's counterargument, however, was cast in a largely philosophical tone.
		\item Laue's response to this issue marked a turning point in the debate, shifting the research focus toward continuum mechanics. Through a straightforward analysis of degrees of freedom, he refuted the definitions of a rigid body\cite{laue1911diskussion}. Furthermore, he investigated the generalization of elasticity theory, including the energy-momentum theorem, within the framework of relativity\cite{laue1911dynamik}.
	\end{itemize}
	\item  1912:  Einstein mentioned the Hamiltonian in the note added in proof to his paper \cite{einstein1912theory}, where he observed that the Hamiltonian is closely related to the spacetime interval. The purpose of Einstein in that article was to make the principle of relativity admit the existence of gravitation, and he pointed out that the Hamiltonian principle `$\cdots$ gives an idea about how the equations of motion of the material point in a dynamic gravitational field are constructed.' Since then, using the variational principle to drive the equation of motion of material particles had become routine practice in his papers, as he repeatedly emphasized this idea\cite{einstein1913outline,einstein1914nor,einstein1914formal}, despite only arriving at the final correct equation of GR in 1914\cite{einstein1914formal}.
	\item   1921: Fermi discuss the motion of an acceleration electron without rotation. Perhaps it was not his original intention, but his discussion provided a way to describe a single observer\cite{fermi1921sulla}.
	\item   1923: 
	\begin{itemize}
		\item Eddington explicitly recognized that not all observers can cover the entire spacetime manifold \cite{eddington1923mathematical}. The term `horizon' had already appeared in the context of his analysis of the de Sitter spacetime. 
		\item  E. Cartan introduced the method of moving frames to describe observers, aiming to clarify concepts such as parallel transport, the equivalence principle, and the role of observers within the framework of general relativity \cite{cartan1923varietes}.
	\end{itemize}
	\item   1935: Walker explicitly connected Fermi's work on the transport of frames to the description of a single observer, introducing what he termed the `non-rotating frame of reference' associated with any observer \cite{walker1935note}.
	\item1942-1952:  Discussions on rigidity gradually diminished, with subsequent efforts largely aimed at clarifying earlier concepts \cite{berenda1942problem,rosen1947notes,fokker1949space,gardner1952rigid}. The definition adopted above, for instance, draws upon the work of Rosen \cite{rosen1947notes}.
	\item  1956-1958:  Rindler, from the perspective of the observer, clarified to some extent the confusion that had arisen in the works of others before him\cite{rindler1956visual}. Combined with the investigations of singularities (both coordinate and intrinsic) by Synge \cite{synge1950gravitational} and Finkelstein \cite{finkelstein1958past}, the work laid the groundwork for subsequent advances in black hole physics.
\end{itemize}

\subsection{Theoretical Reflections}

Through the lens of historical investigation, it can be seen that the development of relativistic mechanics is marked by both coincidence and remarkable ingenuity. The understanding of observers is also gradually deepening in development.

Starting with the description of the motion of a particle, one can see that Planck used analytical mechanics to discuss dynamical problems, promoting the generalization of relativistic mechanics. A fortunate aspect of this approach is that the observers' description does not affect the variational principle, allowing the discussion to proceed smoothly.

Then, turning to the discussion on rigidity, the precise definition of rigidity has become a controversial issue, highlighting the role of observers and leading to the establishment of relativistic hydrodynamics and theory of elasticity. The discussion of Laue provides a mathematical framework for describing the continuum.

Einstein's contributions to the development of GR are often regarded as a quintessential example of individual heroism, while also being seen as supporting Kuhn's perspective that GR represents a paradigm shift in scientific thought. However, it is unreasonable to view this paradigm shift as arising without any antecedents. Pais provides a remarkable narrative of Einstein's intellectual journey toward the formulation of GR\cite{pais1982subtle}, from Einstein's initial realization in 1907 of the connection between gravitation and accelerated motion, to his deep confusion in 1914 regarding how to formulate a gravitational field equation consistent with the equivalence principle, and finally to his discovery of the correct framework. However, the present discussion may offer a different perspective. It is well known that Einstein finally reached GR  at the end of 1915, but there hasn't been much discussion on how he switched from particle-based framework to a field-theoretic formalism\cite{cao2019conceptual}. A closer look at the evolution of relativistic mechanics, especially from the observer's perspective, the transition in Einstein's thinking had already started to manifest itself at that stage.

At least four events occurred during this period that greatly promoted the establishment of the theory: 

\begin{enumerate} 
\item Einstein found the dynamical equation of a particle in relativity can be deduced from a purely  geometric  quantity \cite{einstein1912theory}, the spacetime interval $\D s$. It serves as a guiding principle, combined with the principle of equivalence, and later became the source for deriving the equations of particle motion in GR. 
    \item  It can be found in Einstein's manuscript during 1912-1914 \cite{einstein1912manuscript}, he paid a lot of effort to study the `world geometry' language invented by Minkowski and developed by Abraham \cite{abraham1909electrodynamics}, Sommerfeld \cite{sommerfeld1910relativitatstheorie}, and Laue \cite{laue1911dynamik}. Especially, there is a large section discussing the energy-momentum tensor. Given that prior to 1912, Einstein had discussed relativistic transformations in thermodynamics but never rigorously addressed the continuum mechanics, the discussion of the energy-momentum tensor in his manuscript indicates that he had adopted and was proficient in Laue's discussions. 
        \item Einstein mentioned Mach's idea more than once when discussing the relationship between acceleration and gravitation, but Newton's bucket argument was first mentioned in 1914 \cite{einstein1914formal}. This scenario is highly similar to the one discussed in the Ehrenfest paradox, but it seems that not many people have paid attention to the role of acceleration in it before. 
            \item In his letter to Ehrenfest on 26 December 1915, he wrote,`$\cdots$ the reference system has no real meaning... Whatever is physically real in events in the universe (as opposed to that which is dependent on the choice of a reference system) consist in spatio-temporal coincidences and in nothing else.' From this, it is clear that he had already made a distinction between geometric objects and certain arbitrarily chosen descriptive representations. 
\end{enumerate} 

`Why were another seven years required for the construction of the general theory of relativity? The main reason lies in the fact that it is not so easy to free oneself from the idea that coordinates must have an immediate metrical meaning.'This is a frequently quoted paragraph in Einstein's autobiography\cite{einstein1949albert}. However, he did not further elaborate on the origins of the idea that the coordinate description is independent of the geometric object itself. Based on the historical investigation presented in this paper, it is reasonable to draw the following conclusion: discussions on rigidity and rotation likely played a significant role in shaping his ideas to some extent. J. Stachel has also expressed similar ideas in his work\cite{stachel1980rigidly}. Of course, the deep connection between acceleration and gravitation was only observed by him alone, and he developed the whole theory based on it. However, the related discussions during the same period that provided Einstein with new ideas and tools should not be underestimated.

After the establishment of GR, the understanding of the concept of observers continued to play a crucial role. For example, The reinvestigation of the horizon and the discussion on congruences of timelike and lightlike curves inspiring the Renaissance of Relativistic are both inextricably linked to the observers' description. A physics result that is closer to us in time is Hawking radiation\cite{hawking1975particle}, {\it i.e.} the observers at asymptotic infinity will see particles emitted from the vacuum defined by the observers at the horizon. Roughly speaking, it is resulted in the distinct definitions of the vacuum state adopted by different observers in curved spacetime. From this point, black holes can be regarded as a thermodynamic system, and the related discussions continue to this day. 

Today, we are witnessing a growing number of articles discussing observer dependence\cite{kichenassamy2023relativistic,de2025gravitational,harlow2026quantum,abdalla2025gravitational}. Perhaps further exploration in this direction could aid in deepening our understanding of some more fundamental physical problems.

\section{Conclusion}
\label{sec5}

Historical research makes isolated events appear continuous and smooth. Through historical exploration, we see that the observer's perspective has almost always accompanied the development of relativity. By revisiting some previously controversial issues, we have demonstrated how the understanding of observers in theory of relativity has evolved to today.  However, the trajectory of history is not linear. For the sake of clarity, we have inevitably overlooked contributions of many individuals along this developmental path. Moreover, certain understandings that may now seem incorrect, such as many paradoxes in relativity,  have played a significant role in advancing theoretical progress. 

Throughout history, the observer dependence of phenomena and the invariance of laws in relativity have shaped the dominant modes of thought in different periods. While most of the historical facts presented in this study are not new, certain aspects appear to have been previously overlooked in the scholarly literature. For instance, through a historically grounded narrative, this article demonstrates  that mechanical phenomena involving observer dependence can in fact be formulated within a covariant framework.  Perhaps the most significant contribution of this work is the  clarification of two key points. First, why early relativity scholars did not distinguish between coordinate systems and reference frames, yet still obtained meaningful results? This fact is illustrated through an example that uses the variational method to derive equations of motion after introducing the description of observers. Second, it is noted that Einstein initially paid limited attention to Planck's analytical mechanics formulation in SR. Although he was certainly aware of the broader significance of analytical mechanics, it was only after Einstein discovered the principle guiding the construction of the action that his thinking began to shift, and this also paved the way towards GR.

Furthermore, 
this approach
may also be useful for studying the historical development of other problems in the relativity.

\appendix

\section{Covariance of Euler-Lagrange equation}
\label{appA}

The state space of particle in Lagrangian formalism is the tangent bundle $T\mathcal{M}$ over spacetime $\mathcal{M}$, so we can check covariance of Euler-Lagrange equation \eqref{EL-ab} on it. Suppose that $(x^{\mu},\dot{x}_{t}^{\mu})$ is a series coordinates on $T\mathcal{M}$ with coordinate basis $\{\partial/\partial x^{\mu},\partial/\partial \dot{x}^{\mu}\}$. Then the coordinate transformation $x^{\mu}\rightarrow y^{\rho}(x)$ on spacetime $\mathcal{M}$   induces  a coordinate transformation on $T\mathcal{M}$,
\begin{align}
	(x^{\mu},\dot{x}_{t}^{\mu})\quad\longrightarrow\quad\(y^{\rho}(x),\dot{y}^{\rho}\),
\end{align}
where $\dot{y}^{\rho}\equiv\dot{x}_{t}^{\mu}\dfrac{\partial y^{\rho}}{\partial x^{\mu}}(x)$. Substituting the coordinate transformation into eq.~\eqref{EL-ab}, we obtain
\begin{align}
	&\dfrac{\D }{\D t}\(\dfrac{\partial\mathscr{L}}{\partial \dot{x}_{t}^{\mu}}\)-\dfrac{\partial \mathscr{L}}{\partial x^{\mu}}\notag\\
	=&\dfrac{\D }{\D t}\[\(\dfrac{\partial y^{\rho}}{\partial\dot{x}_{t}^{\mu}}\dfrac{\partial}{\partial y^{\rho}}+\dfrac{\partial \dot{y}^{\rho}}{\partial\dot{x}_{t}^{\mu}}\dfrac{\partial}{\partial \dot{y}^{\rho}}\)\mathscr{L}\]-\(\dfrac{\partial y^{\rho}}{\partial x^{\mu}}\dfrac{\partial}{\partial y^{\rho}}+\dfrac{\partial \dot{y}^{\rho}}{\partial x^{\mu}}\dfrac{\partial}{\partial \dot{y}^{\rho}}\)\mathscr{L}\notag\\
	=&\dfrac{\D }{\D t}\(\dfrac{\partial y^{\rho}}{\partial x^{\mu}}\dfrac{\partial \mathscr{L}}{\partial \dot{y}^{\rho}}\)-\dfrac{\partial y^{\rho}}{\partial x^{\mu}}\dfrac{\partial\mathscr{L}}{\partial y^{\rho}}-\dot{x}_{t}^{\nu}\dfrac{\partial^{2} y^{\rho}}{\partial x^{\mu}\partial x^{\nu}}\dfrac{\partial\mathscr{L}}{\partial \dot{y}^{\rho}}\notag\\
	=&\dfrac{\partial y^{\rho}}{\partial x^{\mu}}\[\dfrac{\D }{\D t}\(\dfrac{\partial \mathscr{L}}{\partial \dot{y}^{\rho}}\)-\dfrac{\partial\mathscr{L}}{\partial y^{\rho}}\]+\dfrac{\partial \mathscr{L}}{\partial \dot{y}^{\rho}}\dfrac{\D }{\D t}\dfrac{\partial y^{\rho}}{\partial x^{\mu}}-\dot{x}_{t}^{\nu}\dfrac{\partial^{2} y^{\rho}}{\partial x^{\mu}\partial x^{\nu}}\dfrac{\partial\mathscr{L}}{\partial \dot{y}^{\rho}}\notag\\
	=&\dfrac{\partial y^{\rho}}{\partial x^{\mu}}\[\dfrac{\D }{\D t}\(\dfrac{\partial \mathscr{L}}{\partial \dot{y}^{\rho}}\)-\dfrac{\partial\mathscr{L}}{\partial y^{\rho}}\],
\end{align}
which shows that Euler-Lagrange equation is a covariant vector equation.

\clearpage 





\section*{Acknowledgments}

This work was supported by National Natural Science Foundation of China (Grant No. T2241005).

\bibliographystyle{cas-model2-names}

\bibliography{cas-refs}



\end{document}